\documentclass[fleqn,twoside]{article}
\usepackage[headings]{espcrc2}

\usepackage{graphicx}

\newcommand{\apm}[2]{\ensuremath{^{+#1}_{-#2}}}
\newcommand{\MeV}{\mathrm{Me\kern -0.1em V}}
\newcommand{\GeV}{\mathrm{Ge\kern -0.1em V}}
\newcommand{\TeV}{\mathrm{Te\kern -0.1em V}}

\newcommand{\kg}{\ensuremath{\kappa_\gamma}}
\newcommand{\kZ}{\ensuremath{\kappa_{\mathrm{Z}}}}
\newcommand{\Lg}{\ensuremath{\lambda_\gamma}}
\newcommand{\LZ}{\ensuremath{\lambda_{\mathrm{Z}}}}
\newcommand{\giZ}{\ensuremath{g_1^{\mathrm{Z}}}}

\newcommand{\WW}{\ensuremath{\mathrm{WW}}}
\newcommand{\EE}{\ensuremath{\mathrm{e}^+\mathrm{e}^-}}
\newcommand{\EN}{\ensuremath{\mathrm{e}\nu}}
\newcommand{\PW}{\ensuremath{\mathrm{W}}}
\newcommand{\EEWW}{\EE\ensuremath{\rightarrow}\WW}
\newcommand{\RS}{\ensuremath{\sqrt{s}}}
\newcommand{\RA}{\ensuremath{\rightarrow}}
\newcommand{\etal}{{\it et al.}}

\newcommand{\NIM}    {Nucl. Inst. Meth.\ }
\newcommand{\NP}     {Nucl. Phys.\ }
\newcommand{\PL}     {Phys. Lett.\ }
\newcommand{\PR}     {Phys. Rev.\ }
\newcommand{\PRL}    {Phys. Rev. Lett.\ }

\title{Gauge boson couplings at LEP}

\author{S. Villa\address[EPFL]{Swiss Federal Institute of Technology, 
        EPFL, LPHE-IPEP, 1015 Lausanne, Switzerland}}%

\runtitle{Gauge boson couplings at LEP}
\runauthor{S. Villa}

\begin{document}

\begin{abstract}
A review is given of the measurements of triple and quartic 
couplings among the electroweak gauge bosons
performed at LEP by the four experiments ALEPH, DELPHI, L3 and OPAL.
Emphasis is placed on recently published results and on combinations of 
results performed by the LEP electroweak gauge-couplings group.
All measurements presented are consistent with the Standard Model expectations.

\vspace{1pc}
\end{abstract}

\maketitle

\section{INTRODUCTION}
Self couplings among electroweak gauge bosons are a consequence of the non-Abelian 
structure of the Standard Model (SM) of electroweak interactions~\cite{theSM}.
The measurement of triple (TGC) and quartic couplings (QGC) 
has been one of the main activities
of the LEP experiments (ALEPH, DELPHI, L3 and OPAL) after the 
\EE \ centre-of-mass energy (\RS) was raised 
above the threshold for the production of two W bosons (LEP2 phase). 
Experiments have performed both precision measurements of couplings as 
predicted by the theory and searches for anomalous couplings which could be
present as a consequence of new physics.
The theoretical input is normally expressed in the form of an effective 
Lagrangian containing all operators of a given order which contribute to 
the set of vertices under study and which
are consistent with basic physical assumptions such as Lorentz 
and electromagnetic gauge invariance.
The extraction of couplings is then performed either by fitting distributions of 
a chosen set of sensitive variables or by means of optimal observables 
techniques~\cite{optimalobs}, in which the sensitivity to the couplings 
is concentrated in a reduced number of properly defined variables.  

All results reported in this paper are based on data collected at LEP during  
1996--2000 at $161\ \GeV < \RS < 209 \ \GeV$, corresponding to
total integrated luminosities up to about 700 $\mathrm{pb}^{-1}$ per experiment. 
For many channels and couplings, LEP combined results exist. The latest combinations
are summarized in~\cite{LEPEWWGGC} where reference is given to the
individual publications. More recent results are not yet combined and
will therefore be referenced directly in the following.
The individual references should be consulted for details about 
experimental techniques and composition of data sets. 

This document is organized as follows: Section~\ref{secCTGC} deals with charged triple 
couplings, {\it i.e.} couplings between the W and neutral 
bosons (WWZ and WW$\gamma$ vertices), Section~\ref{secNTGC}
describes searches for purely neutral TGCs, not foreseen in the SM, and finally
Section~\ref{secQGC} reports limits on QGCs.

\section{CHARGED TRIPLE GAUGE COUPLINGS\label{secCTGC}}
The WWZ and WW$\gamma$ vertices can be probed at LEP2 in WW, single W and 
single $\gamma$ production.
To lowest order, three Feynman graphs contribute to WW production: the s-channel 
Z and $\gamma$ production and the t-channel neutrino exchange. The two s-channel 
diagrams contain the WWZ and WW$\gamma$ vertices.  WW$\gamma$ is also involved
in the single W process ($\EE\rightarrow\PW\EN$) via W-$\gamma$ fusion 
and in single $\gamma$ production ($\rm e^+e^- \rightarrow
\nu_{\mathrm{e}}\bar\nu_{\mathrm{e}}\gamma$), 
via W-boson fusion. 
The most general parametrization of the WWZ and WW$\gamma$ vertices can be
written as a function of seven complex TGCs for each vertex~\cite{hagiwara}. 
Some assumptions are normally made in order to reduce the number of 
parameters to be measured.
Imposing the conservation of C and P symmetry and electromagnetic gauge 
invariance reduces the number
of free parameters to five, usually chosen to be \giZ, \kg, \Lg, \kZ \ and
\LZ. At tree level within the SM, $\giZ = \kg = \kZ = 1$ and $\Lg = \LZ =1$.
A further reduction is normally achieved by the requirement of custodial $\mathrm{SU(2)}$
gauge invariance, which translates to the constraints 
$\kZ = \giZ - \tan^2\theta_{W} (\kg-1)$ and $\LZ =
\Lg$~\cite{DKLDXYSchild,LEP2YRAC}, where $\theta_{W}$
is the weak mixing angle. 
\giZ, \kg \ and \Lg \ are normally chosen as the minimal set of parameters to be
measured and are the ones used in the LEP-wide combinations.

\begin{figure}[htb]
\includegraphics[width=15pc]{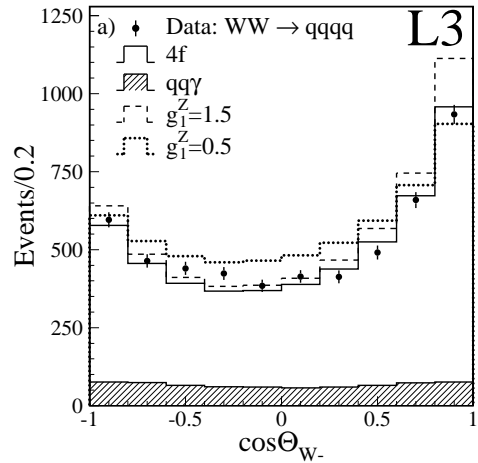}
\caption{Distribution of the W$^{-}$ production angle in hadronic
W-pair events collected by the L3 detector. 
Data are shown together with the expectations for
the SM and for anomalous values of \giZ.}
\label{fig:qqqqcosw}
\end{figure}

The WW channel gives the best sensitivity at LEP in the measurement of charged TGCs. 
Anomalous couplings would affect the total production cross section, the production angle
and the polarization of the W bosons. 
A WW event, yielding two fermions and two anti-fermions in the
final state,  is completely described in terms of five 
angles: the W production angle and the polar and azimuthal angles of the fermion
in the W$^{-}$ rest frame and of the anti-fermion in the W$^{+}$ rest frame.
The distributions of these angles, or other quantities derived from them, are 
used by the LEP experiments to derive measurements of TGCs.
As an example, the distribution of the cosine of the W$^{-}$ production angle in
fully hadronic WW events collected by L3~\cite{l3tgc} 
is shown in Figure~\ref{fig:qqqqcosw}, where the effect
of anomalous values of \giZ \ is also plotted.

\begin{figure}[htb]
\includegraphics[width=18pc]{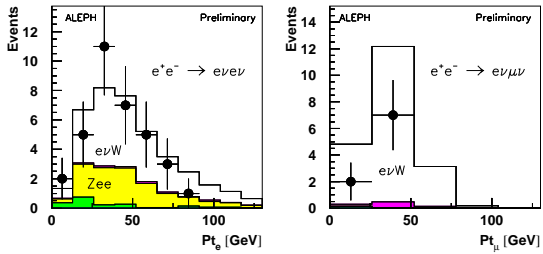}
\caption{Distributions of the transverse momentum of electrons (left) 
and muons (right) in single W events collected by ALEPH.}
\label{fig:singleW}
\end{figure}

The single W channel gives a sensitivity comparable to that obtained
with WW events only on \kg; on the other hand it gives the possibility to
disentangle the effects of the WWZ and WW$\gamma$ vertices, which are 
correlated in measurements based on the WW channel.
In this process, the electron is produced at very low polar angle and
is therefore not detected, leaving a signature of only one W boson 
in the detector.
LEP experiments use different kinematic variables to extract the
couplings from the single W channel. As an example the transverse momentum 
of the electrons and muons in leptonically decaying W's as measured by
ALEPH are shown in Figure~\ref{fig:singleW}.
The energy spectrum and the angular distribution of the photon produced in 
single $\gamma$ processes are sensitive to the WW$\gamma$ vertex, even though
less than the WW and single W channels. 

\begin{figure}[htb]
\includegraphics[width=19pc]{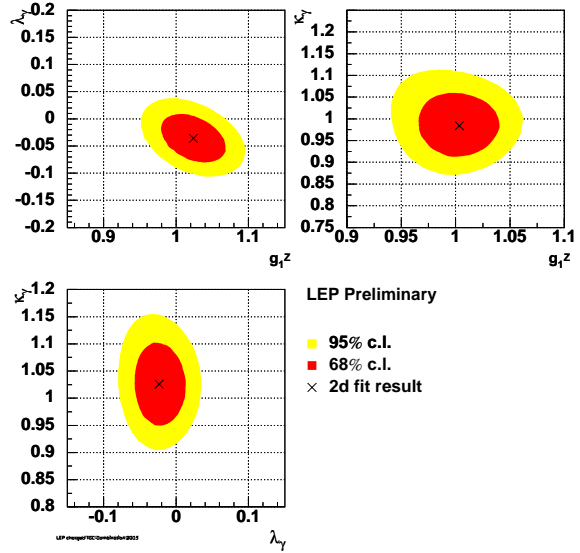}
\caption{The 68\% C.L. and 95\% C.L. contours for two-parameter fits of charged TGCs 
(LEP combination). In each case, the third parameter is set to 
its SM value and the constraints 
$\kZ = \giZ - \tan^2\theta_{W} (\kg-1)$ and $\LZ = \Lg$ 
are imposed.}
\label{fig:tgc2d}
\end{figure}

Measurements of charged TGCs by the four 
LEP experiments are summarized in Table~\ref{table:tgc}, together with 
their combination~\cite{LEPEWWGGC}. All results are obtained by varying one
parameter while setting the others to their SM value. 
The constraints $\kZ = \giZ - \tan^2\theta_{W} (\kg-1)$ and $\LZ = \Lg$ 
are imposed.
\begin{table*}[htb]
\caption{Results of the measurements of charged TGCs by the LEP experiments
and their combinations. Statistical and systematic uncertainties are
included.}
\label{table:tgc}
\newcommand{\m}{\hphantom{$-$}}
\newcommand{\cc}[1]{\multicolumn{1}{c}{#1}}
\renewcommand{\tabcolsep}{1pc} 
\renewcommand{\arraystretch}{1.2} 
\begin{tabular}{@{}llllll}
\hline
 parameter  &  \cc{ALEPH}   &   \cc{DELPHI}  &    \cc{L3}    & \cc{OPAL}    &   \cc{LEP}   \\
\hline
\giZ     &1.026\apm{0.034}{0.033}&1.002\apm{0.038}{0.040}&
          \m 0.928\apm{0.042}{0.041}&\m 0.985\apm{0.035}{0.034}&\m 0.991\apm{0.022}{0.021} \\
\kg      &1.022\apm{0.073}{0.072}&0.955\apm{0.090}{0.086}&
          \m 0.922\apm{0.071}{0.069}&\m 0.929\apm{0.085}{0.081}&\m 0.984\apm{0.042}{0.047} \\
\Lg      &0.012\apm{0.033}{0.032}&0.014\apm{0.044}{0.042}&
          $-0.058$\apm{0.047}{0.044}&$-0.063$\apm{0.036}{0.036}&$-0.016$\apm{0.021}{0.023} \\
\hline
\end{tabular}
\end{table*}
The LEP combined results of fits to two of the three TGCs are shown in Figure~\ref{fig:tgc2d}.
All measurements are consistent with the SM expectations.

\section{NEUTRAL TRIPLE GAUGE COUPLINGS\label{secNTGC}}
Triple couplings among neutral electroweak gauge bosons are not present in the 
SM at tree level, and get contributions of the order of $10^{-4}$ 
from loop diagrams. Such couplings could be enhanced by new physics  
effects~\cite{gounaris2000} and it is therefore important to set limits on their strength.
The $\gamma\gamma$Z, $\gamma$ZZ and ZZZ vertices can 
give as on-shell final state bosons either  Z$\gamma$ or ZZ.
Two independent parametrizations describe the two cases: the eight 
$h_{i}^\mathrm{V}$ couplings ($i=1,\ldots,4$)   for
the Z$\gamma$ final state and the four $f_{i}^\mathrm{V}$ couplings 
($i=4,5$) for the ZZ final state~\cite{hagiwara,gounaris2000}, 
where in both sets V stands for the s-channel exchanged off-shell boson,
which can be a Z or a $\gamma$.

The Z$\gamma$ final state is produced at LEP by events in 
which the photon is radiated off the incoming 
electron or positron. 
The presence of anomalous couplings would increase the total cross section and 
enhance the photon production at large polar angles. 
By fitting distributions of kinematic variables such as the photon energy, 
polar angle and recoil mass, experiments are able to set limits on the
$h_{i}^\mathrm{V}$ couplings. 
The LEP combined results~\cite{LEPEWWGGC} from fits where
each parameter is measured separately while setting all others
to zero, are summarized in Table~\ref{table:ntgc}
as 95\% confidence level (C.L.) limits. 
\begin{table}[htb]
\caption{Results of the combination of the measurements of neutral TGCs 
by the LEP experiments. Statistical and systematic uncertainties are
included.}
\label{table:ntgc}
\newcommand{\m}{\hphantom{$-$}}
\newcommand{\z}{\phantom{0}}
\newcommand{\cc}[1]{\multicolumn{1}{c}{#1}}
\renewcommand{\tabcolsep}{2pc} 
\renewcommand{\arraystretch}{1.2} 
\begin{tabular}{@{}cc}
\hline
parameter  &  \cc{95\% C.L.}  \\
\hline
$h_{1}^{\gamma}$   	  & $[-0.056,+0.055]$	 \\
$h_{2}^{\gamma}$   	  & $[-0.045,+0.025]$	 \\
$h_{3}^{\gamma}$   	  & $[-0.049,-0.008]$	 \\
$h_{4}^{\gamma}$   	  & $[-0.002,+0.034]$	 \\
$h_{1}^\mathrm{Z}$	  & $[-0.13\z,+0.13\z]$  \\
$h_{2}^\mathrm{Z}$	  & $[-0.078,+0.071]$	 \\
$h_{3}^\mathrm{Z}$	  & $[-0.20\z,+0.07\z]$  \\
$h_{4}^\mathrm{Z}$	  & $[-0.05\z,+0.12\z]$  \\[3pt]
$f_{4}^{\gamma}$          & $[-0.17\z,+0.19\z]$  \\
$f_{4}^\mathrm{Z}$	  & $[-0.30\z,+0.30\z]$  \\
$f_{5}^{\gamma}$          & $[-0.32\z,+0.36\z]$  \\
$f_{5}^\mathrm{Z}$	  & $[-0.34\z,+0.38\z]$  \\
\hline
\end{tabular}
\end{table}
Final updated results have been published by L3~\cite{L3Zgamma}, which are not 
taken into account in this combination.

\begin{figure}[htb]
\includegraphics[width=16pc]{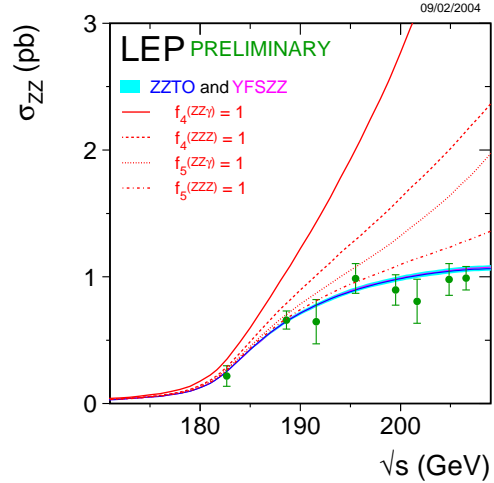}
\caption{The LEP combined $\EE\rightarrow\mathrm{ZZ}$ cross section. SM prediction
and data points are shown together with curves corresponding to anomalous 
values of the $f_{4,5}^{\mathrm{V}}$ couplings.
}
\label{fig:ZZxsec}
\end{figure}

The ZZ final state is accessible at LEP2 via t-channel electron exchange. An
s-channel $\gamma^{*}/\mathrm{Z}^{*}$ would yield the same final state if
neutral $f_{4,5}^{\mathrm{V}}$ TGCs were present. 
These anomalous couplings would increase the total cross section and modify
the event kinematics.
ZZ production has been studied at LEP in all of its visible topologies, 
and no deviations from the SM predictions have been observed.
The measured ZZ production cross section as a function of \RS \ is plotted in
Figure~\ref{fig:ZZxsec}, where the effect of non-zero anomalous couplings is 
also shown.
Combined LEP results on the $f_{4,5}^{\mathrm{V}}$ couplings are summarized 
in Table~\ref{table:ntgc} in the form of 95\% C.L. intervals.

\section{QUARTIC GAUGE COUPLINGS\label{secQGC}}
Quartic couplings are present in the SM Lagrangian, but their size is 
expected to be too small to be measurable at LEP.
The LEP experiments have therefore performed measurements to set limits
on the existence of anomalous QGCs, both of the type predicted by the electroweak 
theory and of more general form, which might be induced by new physics.
The parametrization used~\cite{QGCtheory} only considers quartic terms in the Lagrangian which
are not associated with TGCs, since the other terms are more strongly constrained 
by the measurements discussed in Section~\ref{secCTGC} and \ref{secNTGC}.
Given the kinematic limit of LEP (maximum of two on-shell heavy bosons produced), 
the following vertices can be tested: WW$\gamma\gamma$, parametrized by
$a_{0}^{\mathrm{W}}$ and $a_{c}^{\mathrm{W}}$,
WWZ$\gamma$, parametrized by $a_{n}$, and ZZ$\gamma\gamma$ parametrized by
$a_{0}^{\mathrm{Z}}$ and $a_{c}^{\mathrm{Z}}$.
Only the first two vertices are present in the SM. Due to the dimension 
of Lagrangian operators containing four gauge bosons, quartic couplings have 
dimensions of the inverse of a squared energy and are therefore 
normally written as proportional to $1/\Lambda^{2}$, where $\Lambda$ is
the typical energy scale of the new physics responsible for the QGCs.

\begin{figure}[htb]
\includegraphics[width=18pc]{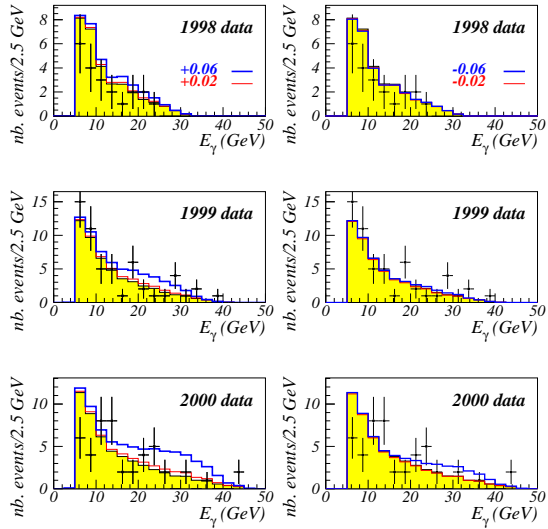}
\caption{Photon energy spectrum in the $\EEWW\gamma$ channel as measured
by DELPHI. The three data sets correspond to $\RS= 189\ \GeV$ (top), $\RS= 198\ \GeV$ (middle)
and $\RS= 206\ \GeV$ (bottom). Data (dots) are compared to SM expectation (shaded histograms)
and to the corresponding distributions for positive (left) and negative (right) values of 
$a_{c}^{\mathrm{W}}$ (in $\GeV^{-2}$).}
\label{fig:WWgamma}
\end{figure}

At LEP, three topologies can give information about QGCs. The first
one is $\EEWW\gamma$, which is sensitive to the couplings  
$a_{0}^{\mathrm{W}}$, $a_{c}^{\mathrm{W}}$ and
$a_{n}$. Events with this topology are produced 
via radiative W-pair production, where a photon is radiated from one of the initial
or final state fermions or from one of the W bosons.
Anomalous couplings would increase the cross section and modify the photon 
energy spectrum, as can be seen in Figure~\ref{fig:WWgamma}, where DELPHI 
data~\cite{DELPHIWWgamma} is
shown together with Monte Carlo expectations for SM and for some non-zero
values of the coupling $a_{c}^{\mathrm{W}}$.

\begin{figure}[htb]
\includegraphics[width=18pc]{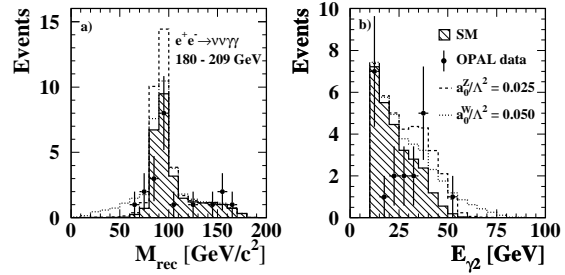}
\caption{Recoil mass (left) and energy of the second highest energetic photon (right)
for $\nu\bar{\nu}\gamma\gamma$ events collected by 
OPAL. Data and SM expectations are shown, together with
predictions for non-zero values of $a_{0}^{\mathrm{Z}}$
and $a_{0}^{\mathrm{W}}$.}
\label{fig:opalnng}
\end{figure}

The second channel sensitive to QGCs is $\EE\RA\nu\bar{\nu}\gamma\gamma$. This
final state can be produced both through W fusion into a WW$\gamma\gamma$ vertex
and via s-channel Z production, a ZZ$\gamma\gamma$ vertex and subsequent decay
of the Z into a neutrino pair. The same signature of two acoplanar photons is 
used in separate fits to determine the W and Z couplings, each time varying 
only one parameter while setting 
all others to zero.  Kinematic variables 
normally used in the fit are the recoil mass
and the energy of the second highest energetic photon. 
In Figure~\ref{fig:opalnng}, distributions of these variables 
as measured by OPAL~\cite{OPALnngg} are plotted together with expectations
for SM and for anomalous values of the parameters $a_{0}^{\mathrm{Z}}$
and $a_{0}^{\mathrm{W}}$. 
The ALEPH collaboration
has also recently published~\cite{ALEPHnngg} limits on QGCs extracted from this channel.

\begin{figure}[htb]
\includegraphics[width=14pc]{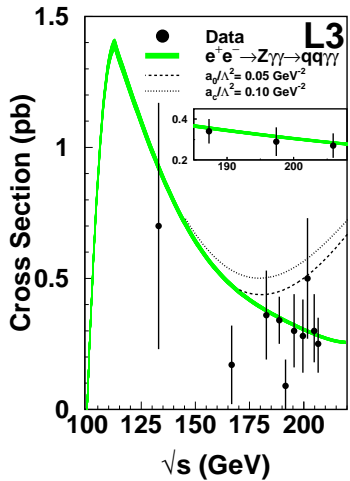}
\caption{The L3 measurement of the cross section of the process 
$\EE\RA\mathrm{Z}\gamma\gamma \RA \mathrm{q\bar{q}}\gamma\gamma$ as a function
of \RS. Dashed and dotted lines represent anomalous QGC predictions. The inset
presents the same results with the highest energy data combined in three 
samples.}
\label{fig:zggL3}
\end{figure}

The $\EE\RA\mathrm{Z}\gamma\gamma \RA \mathrm{q\bar{q}}\gamma\gamma$ channel is
also used at LEP to set limits on the $a_{0}^{\mathrm{Z}}$ and $a_{c}^{\mathrm{Z}}$
parameters. The corresponding SM background is the double radiative return
to the Z. Quartic couplings of the ZZ$\gamma\gamma$ type would  enhance the
total cross section at high \RS \ and modify the photons spectra. 
Figure~\ref{fig:zggL3} shows
the L3 measurement of the $\mathrm{Z}\gamma\gamma$ cross section as a function
of centre-of-mass energy and the predicted behaviour in case of anomalous couplings.

The results of the combination of the ALEPH, L3 and OPAL measurements of
the ZZ$\gamma\gamma$ couplings are $-0.029\ \GeV^{-2} < 
a_{c}^{\mathrm{Z}}/\Lambda^{2} < +0.039\ \GeV^{-2}$ and 
$-0.008\ \GeV^{-2} < a_{0}^{\mathrm{Z}}/\Lambda^{2} < +0.021\ \GeV^{-2}$
at 95\% C.L., including statistical and systematic uncertainties.
For the WW$\gamma\gamma$ and WWZ$\gamma$ couplings, no recent LEP 
combined results exist. The results of the single LEP experiments
are reported in Table~\ref{table:wwgg-wwzg}.
\begin{table*}[htb]
\caption{Measurements of $\mathrm{a_{0}^{W}}$, $\mathrm{a_{c}^{W}}$ 
and $\mathrm{a_{n}}$
by the LEP experiments. Limits are expressed in $\GeV^{-2}$ at 95\% C.L.; 
statistical and systematic uncertainties are included.}
\label{table:wwgg-wwzg}
\newcommand{\cc}[1]{\multicolumn{1}{c}{#1}}
\newcommand{\z}{\phantom{0}}
\renewcommand{\tabcolsep}{1.2pc} 
\renewcommand{\arraystretch}{1.2} 
\begin{tabular}{@{}lcccc}
\hline
parameter &          ALEPH  &    DELPHI    &        L3     &        OPAL   \\
\hline
$\mathrm{a_{0}^{W}}/\Lambda^{2}$&$[-0.060,+0.055]$&$[-0.020,+0.020]$&
                                 $[-0.015,+0.015]$&$[-0.020,+0.020]$ \\ 
$\mathrm{a_{c}^{W}}/\Lambda^{2}$&$[-0.099,+0.093]$&$[-0.063,+0.032]$&
                                 $[-0.048,+0.026]$&$[-0.052,+0.037]$ \\
$\mathrm{a_{n}}/\Lambda^{2}$    & -               &$[-0.18\z,+0.14\z]$  &
                                 $[-0.14\z,+0.13\z]$  &$[-0.16\z,+0.15\z]$   \\	      
\hline
\end{tabular}
\end{table*}
%

\section{CONCLUSION\label{secConclude}}
Results of the measurements of triple and quartic gauge boson couplings
by the four LEP experiments are reviewed in this paper, together with
their combinations.
Most results are either published or being finalized for publication.
The measurement of charged TGCs, which are predicted as a fundamental 
consequence of the gauge structure of the electroweak theory, has reached at
LEP the precision of about 2--4\%, with results in agreement with the SM.
Searches for anomalous triple and quartic couplings have been performed by
all LEP experiments in several different final states, 
and limits are set with no deviations observed from
the SM predictions.


\end{document}